# Propagation of Pendulum Waves under Deep-Seated Cord Charge Blasting in Blocky Rock Mass

## N. I. Aleksandrova

*Chinakal Institute of Mining, Siberian Branch, Russian Academy of Sciences,*
*Novosibirsk, 630091 Russia*
*e-mail: nialex@misd.ru*

**Abstract**—The object of the numerical study is the travel of pendulum waves in a blocky medium under nonstationary impact of deep-seated charge blasting on the surface of the expansion chamber. The blocky model is simulated by a two-dimensional lattice of masses connected by elastic springs along the axes and diagonals. The displacements and velocities of the masses at different half-space points are calculated using the finite-difference method.

## INTRODUCTION

Understanding transient deformation of nonuniform media is required in creating models and methods to calculate and prove estimates of failure of ground and underground structures under seismic waves of earthquakes, mine explosions and vibration treatments.

The recently accomplished researches point at the necessity to include mathematical models of geomechanics and seismicity with the block structure of rocks [1, 2]. A rock mass is considered in this case as a system of nested blocks of different scales with in-between layers composed of weaker jointed rocks. The dynamic behavior of a block-structure medium can be approximately described as movement of rigid blocks owing to pliancy of interlayer [2]. The analytical model in this case is a lattice of masses connected by springs and dampers. Such models can be two-dimensional [3–6] and three-dimensional [6, 7].

This study uses a two-dimensional model to obtain a finite-difference solution of a dynamic problem on impact of deep-seated cord-like explosion on the surface of the explosion cavity in a block-structure medium.

## 1. PROBLEM FORMULATION

In focus is the load exerted by an expansion center on the surface of a cylindrical cavity in a block half-space. The problem is plane. The cavity generators are parallel to the stress-free surface of the half-space. The block medium is modeled as a uniform 2D lattice of masses connected by springs along axes and diagonals (Fig. 1). Here, $x$, $y$ are the horizontal and vertical axes; $n$, $m$ are the numbers of masses in the directions of the axes $x$, $y$. The free surface relates with the value $m = 0$, the half-space — with $m < 0$. The cavity inside the half-space is modeled as a zone without links. The cavity symmetry center is at the point $O$ at a depth $h + l/2$ ($l$ is the length of the springs in the directions $x$, $y$). The axis $y$ is the axis of symmetry of the problem. The forces of the same amplitude and different orientation are applied at four points on the cavity surface. The time dependence of loading is described by the formula: $Q(t) = Q_0 H(t)\sqrt{2}$, where $H(t)$ is the Heaviside function; $Q_0$ is the load amplitude.

Movement of the block medium inside the half-space is given by the equations:

$$M\ddot{u}_{n,m} = k_1(u_{n+1,m} - 2u_{n,m} + u_{n-1,m}) + k_2(u_{n+1,m+1} + u_{n-1,m-1} + u_{n+1,m-1}$$
$$+ u_{n-1,m+1} - 4u_{n,m})/2 + k_2(v_{n+1,m+1} + v_{n-1,m-1} - v_{n-1,m+1} - v_{n+1,m-1})/2,$$
$$M\ddot{v}_{n,m} = k_1(v_{n,m+1} - 2v_{n,m} + v_{n,m-1}) + k_2(u_{n+1,m+1} + u_{n-1,m-1} - u_{n+1,m-1}$$
$$- u_{n-1,m+1})/2 + k_2(v_{n+1,m+1} + v_{n-1,m-1} + v_{n-1,m+1} + v_{n+1,m-1} - 4v_{n,m})/2.$$

(1)

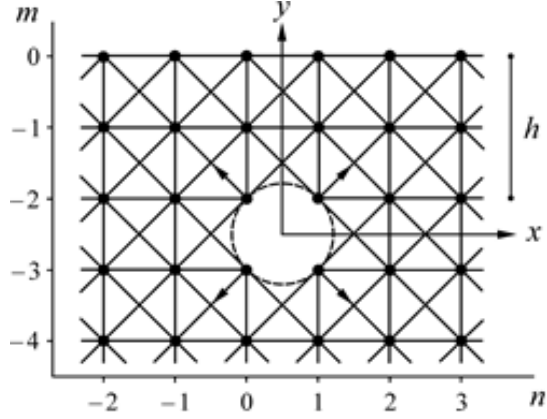

**Fig. 1.** Problem formulation.

Here, $u$, $v$ are the displacements in the directions $x$, $y$; $M$ is the mass of the blocks; $k_1$ are the stiffnesses of the springs in the directions $x$, $y$; $k_2$ are the stiffnesses of the springs in the diagonal directions.

The block movement at the boundary $m = 0$ is described by the equations below:

$$M\ddot{u}_{n,0} = k_1(u_{n+1,0} - 2u_{n,0} + u_{n-1,0}) + k_2(u_{n-1,-1} + u_{n+1,-1} - 2u_{n,0})/2 + k_2(v_{n-1,-1} - v_{n+1,-1})/2,$$
$$M\ddot{v}_{n,0} = k_1(v_{n,-1} - v_{n,0}) + k_2(u_{n-1,-1} - u_{n+1,-1})/2 + k_2(v_{n-1,-1} - 2v_{n,0} + v_{n+1,-1})/2. \qquad (2)$$

The article omits cumbersome equations of the block movement on the cavity surface. The can be written using Fig. 1. The initial conditions are zero.

Let us discuss an isotropic lattice, i.e., when $k_1 = 2k_2$ and long waves travel in all directions at the similar velocities of $P$- and $S$-waves [4]:

$$c_p = l\sqrt{\frac{3k_1}{2M}}, \quad c_s = l\sqrt{\frac{k_1}{2M}}. \qquad (3)$$

According to [5], the velocity of the surface Rayleigh waves is given by:

$$c_R = l\sqrt{\frac{k_1}{M}\left(1 - \frac{1}{\sqrt{3}}\right)}. \qquad (4)$$

## 2. NUMERICAL RESULTS

The equations (1) and (2) were solved using the finite difference method and an explicit scheme. The stability condition in the difference scheme $\tau \leq l\sqrt{2M/3k_1}$ ($\tau$ — is the time step of the difference grid). The problem is symmetrical relative to the axis $y$, and, for this reason, the calculations were only performed in the domain $n \geq 1$ ($x \geq 0$) with the symmetry conditions at the boundary $n = 1$: $u_{0,m} = -u_{1,m}$, $v_{0,m} = v_{1,m}$ ($m \leq 0$).

Figure 2 shows the distribution of the vertical displacement $v$ and its velocity $\dot{v}$ in the half-plane at the time $t = 200$ calculated for two values of the occurrence depth of the plane: $h = 1$ (Fig. 2a) and $h = 25$ (Fig. 2b). Hereinafter, it is assumed that: $M = 1$, $l = 1$, $k_1 = 3/4$, $Q_0 = 1$, $\tau = 0.314$. When the cavity is at the shallow depth, the Rayleigh wave, $R$-wave, forms very soon (Fig. 2a) whereas this process is essentially slow in the deeper-seated cavity (Fig. 2b): the wave is almost unobserved at $t = 200$ ($h = 25$). The $P$-wave emitted by the cavity is reflected from the free surface of the half-space in the form of two waves: longitudinal $PP$-wave and shear $PS$-wave. The fronts of these waves are traced in Fig. 2. The shear wave $S$ emitted by the cavity is also well seen in Fig. 2a.

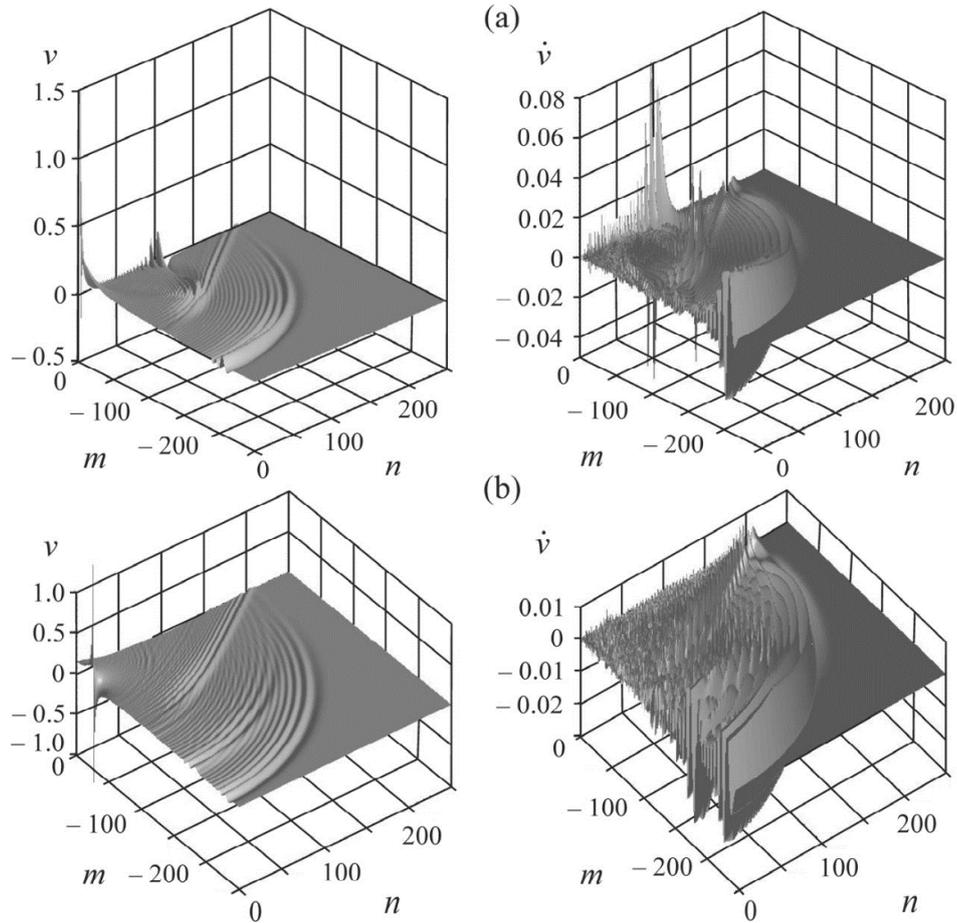

**Fig. 2.** Vertical velocities of displacements in the half-space at the time moment $t = 200$ at the depths $h = 1$ (a) and $h = 25$ (b).

Schematically, the wave fronts are depicted in Fig. 3a, while Fig. 3b shows the calculated vertical displacements at the given depth of the cavity ($h = 66$, $t = 93.3$). The front of $P$-wave at the time $t = 93.3$ is spaced from the source at the distance $r_p = tc_p = 99$. The front of $PS$-wave is an envelope of the fronts of the reflected waves from the half-space surface at the points $(n, 0)$: $r_{ps} = c_s(t - \sqrt{h^2 + (n-1)^2}/c_p)$.

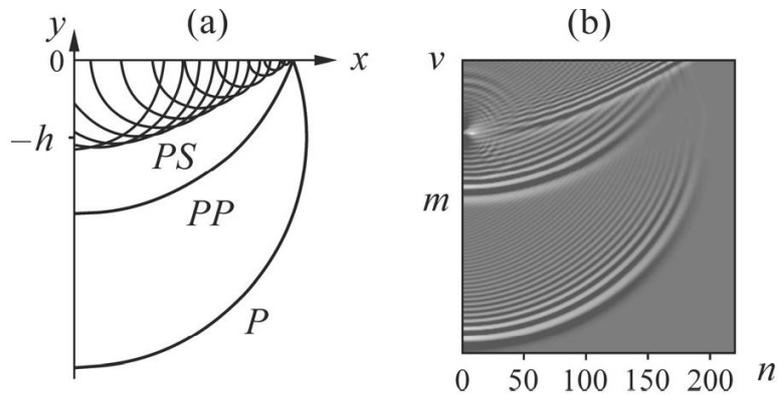

**Fig. 3.** (a) Scheme of the wave fronts and (b) vertical distributions in the plane $(n, m)$ at $h = 66$, $t = 93.3$.

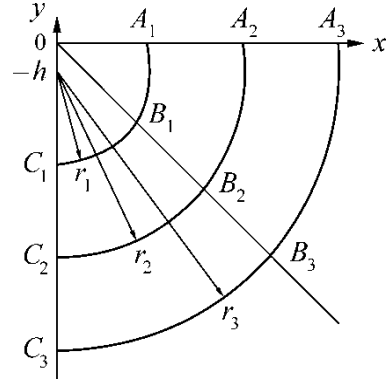

**Fig. 4.** Pattern of the calculation points.

Further on, we study, the influence of the occurrence depth of the cavity on the vertical displacements of blocks at different points of the half-plane (Fig. 4): $A_1, A_2, A_3$ are the points of the half-plane surface, $B_1, B_2, B_3$ are the points on the bisector of the angle $x0y$, $C_1, C_2, C_3$ are the points on the vertical axis $y$. The points having the subscript 1 lie at the distance $r_1 = 100$ from the cavity, the points with the subscript 2 — at $r_2 = 200$, the points with the subscript 3 — at $r_3 = 300$.

For the points $A_i = (n_i, 0)$ $(i = 1, 2, 3)$:

$$t_p = \frac{\sqrt{h^2 + (n_i - 1)^2}}{c_p}$$ — arrival of the front of P-wave,

$$t_R = \frac{h}{c_p} + \frac{n_i - 1}{c_R}$$ — arrival of the front R-wave.

For the points $B_i = (n_i, m_i) = (n_i, 1 - n_i)$, $(i = 1, 2, 3)$, the time of arrival of the front of P-wave is:

$$t_p = \frac{\sqrt{(h - m_i)^2 + (n_i - 1)^2}}{c_p} \qquad \text{when } m_i < -h,$$

$$t_p = \frac{m_i}{c_p} \qquad \text{when } m_i = -h; -h - 1 \text{ and}$$

$$t_p = \frac{\sqrt{(h + 1 - m_i)^2 + (n_i - 1)^2}}{c_p} \qquad \text{when } m_i < -h - 1.$$

The front of PP-wave arrives at $t_{pp} = \sqrt{(h - m_i)^2 + n_i^2}/c_p$; the front of PS-wave — at $t_{ps} = \sqrt{h^2 + x^2}/c_p + \sqrt{(x - n_i)^2 + m_i^2}/c_s$. Here, the unknown variable $x$ is found from the condition that the function $t_{ps}(x)$ reaches a minimum, i.e. $\partial t_{ps}/\partial x = 0$. On this basis, we derive an equation in terms of $x$:

$$x^2[(x - n_i)^2 + m_i^2] - \frac{c_p^2}{c_s^2}(n_i - x)^2(h^2 + x^2) = 0$$

and solve it using a computer.

The calculated arrival times of the wave fronts at the points $C_i = (1, m_i)$ $(i = 1, 2, 3)$ are: $t_p = (h + 1 + m_i)/c_p$ for $P$-wave; $t_{pp} = (h - m_i)/c_p$ for $PP$-wave and $t_{ps} = h/c_p - m_i/c_s$ for $PS$-wave. Similarly $m_i < 0$.

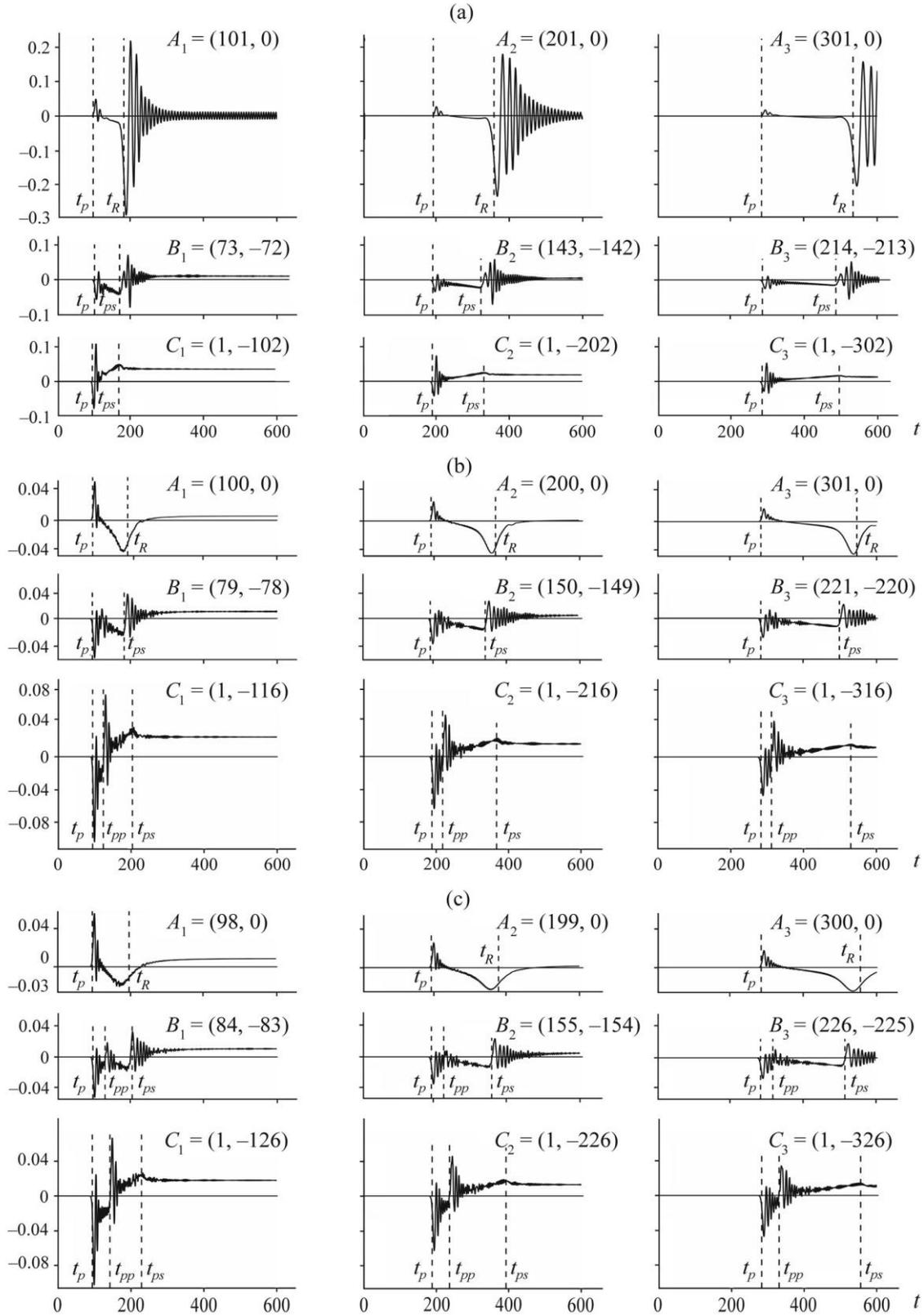

**Fig. 5.** Time dependences of the vertical displacements: (a) $h = 1$; (b) $h = 15$; (c) $h = 25$.

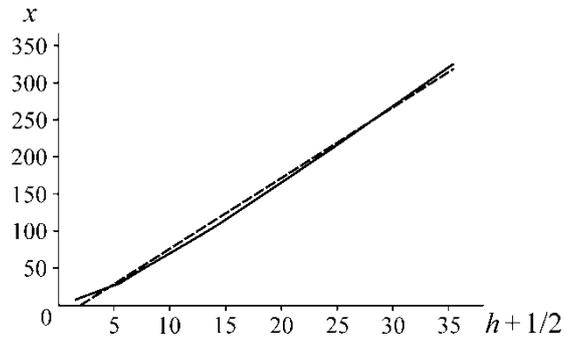

**Fig. 6.** Occurrence depth of the cavity versus of the coordinate of the half-space surface points where the maximum modulus of the amplitude of $v$ in $R$-wave starts to exceed the maximum amplitude of $v$ in $P$-wave.

Figure 5 shows the time dependences of the vertical displacements at the half-plane points under step-wise loading of the cavity: $h=1$, $h=15$, $h=25$. The vertical dashed lines mark the arrival times of fronts of $P$- and $R$-waves and reflected $PP$- and $PS$-waves and are denoted by the related symbol nearby. The coordinates of the calculation points are also given in the figures.

From the analysis of the data in Fig. 5, it follows that at the given $r_i$ ($i=1, 2, 3$), the modulus of the vertical displacements reaches the peak values on the half-plane surface at the points $A_i$ in the vicinity of $R$-wave when the cavity is close-spaced with the surface ($h=1$). The peak modulus of the amplitude of $v$ in the Rayleigh wave is several times higher than the amplitude of $v$ in the longitudinal wave. At the points $C_i$ the modulus of the vertical displacements at any $h$ reaches its maximum in the longitudinal wave. For the points $B_i$ and small $h$, the maximum modulus of the amplitude of $v$ in $PS$-wave can exceed the maximum amplitude $|v|$ in $P$-wave (Fig. 5a); whereas the situation can be otherwise at high values of $h$ (Figs. 5b and 5c).

As the distance $h$ from the half-space surface to the cavity is increase, the maximum values of $|v|$ in the vicinity of the Rayleigh wave front essentially lower and exceed the maximums of $|v|$ in the longitudinal wave at larger distances from the point $x=0$, $y=0$. Figure 6 demonstrates the relationship between the cavity occurrence depth $h+0.5$ and coordinate $x=n-0.5$ of a half-space surface point where the maximum $|v|$ in $R$-wave starts to exceed the maximum amplitude of $v$ in $P$-wave. The solid line shows the calculated results, the dashed line is an approximation of the numerical results: $x=9.555(h+0.5)-20.554$. The coordinate of the half-space surface point at which the maximum modulus of the amplitude of $v$ in the Rayleigh wave starts to exceed the maximum amplitude of $v$ in the longitudinal wave linearly increases with the depth of occurrence of the cavity.

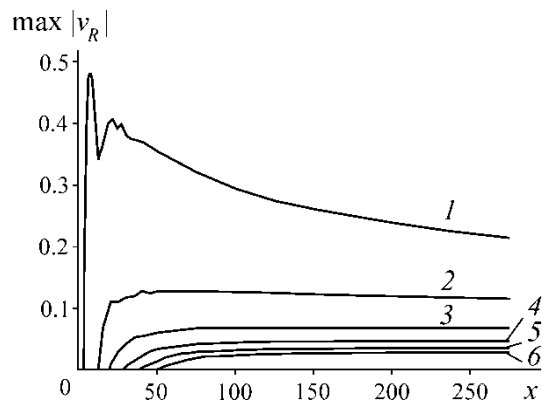

**Fig. 7.** Maximum modulus of the amplitude of $v$ versus $x$. $1 - h=1$; $2 - h=5$; $3 - h=10$; $4 - h=15$; $5 - h=20$; $6 - h=25$.

In case of the relationship between the maximum modulus of the vertical displacement amplitude in the Rayleigh wave, $\max|v_R|$, and the distance $x = n - 0.5$ at different $h$ (Fig. 7), $\max|v_R|$ first grows with the increase in n and then flattens. Furthermore, it is seen in Fig. 7 that $\max|v_R|$ reduces with the deeper seated cavity and it faster reaches a constant value with an increase in $x$.

## CONCLUSIONS

The implemented numerical solution of the plane dynamic problem on the impact exerted by a deep-seated cord-like explosion on the surface of the explosion cavity in a block medium shows that the maximum modulus of the amplitude of the vertical displacements in the Rayleigh wave first grows with an increase in the distance from the cavity, then attenuates and tends to a constant value which lowers with the deeper occurrence of the cavity. The coordinate of the half-space surface point at which the maximum modulus of the amplitude of $v$ in the Rayleigh wave starts to exceed the maximum amplitude of $v$ in the longitudinal wave linearly grows with the occurrence depth of the cavity.

The study shows that the block structure of a medium has an influence on the appearance of high-frequency oscillations in *P*-, *R*-, *PP*- и *PS*-waves.

## ACKNOWLEDGMENTS

This work was supported by the Russian Academy of Sciences, project no. ONZ RAN 3.1.